\documentclass{epl}

\usepackage{amsmath}            
\usepackage{amssymb}
\newcommand{\Leps}{\text{\Large{$ \varepsilon $}}}

\title{Bandgap modulation of narrow-gap carbon nano\-tubes in a transverse electric field}
\shorttitle{Bandgap modulation of carbon nano\-tubes}

\author{D. Gunlycke\inst{1} \and C. J. Lambert\inst{2} \and S. W. D. Bailey\inst{2} \and D. G. Pettifor\inst{1} \and G. A. D. Briggs\inst{1} \and J. H. Jefferson\inst{3}}
\shortauthor{D. Gunlycke \etal}

\institute{                    
  \inst{1} Department of Materials, University of Oxford - Parks Road, Oxford, OX1 3PH, UK.\\
  \inst{2} Department of Physics, University of Lancaster - Lancaster, LA1 4YB, UK.\\
  \inst{3} Sensors and Electronics Division, QinetiQ - St. Andrews Road, Malvern, WR14 3PS, UK.
}

\begin{document}

\maketitle

\begin{abstract}
We propose a method to modulate the bandgaps in narrow-gap carbon nano\-tubes using a transverse electric field. Unlike previous investigations, we include curvature effects of the nano\-tubes by incorporating both $\pi$- and $\sigma$-orbitals in our tight-binding calculations. The calculations show that the narrow curvature-induced bandgaps decrease quadratically with electric field amplitude to zero. As the electric field amplitude continues to increase, the bandgap then expands in a similar manner to that presented in earlier studies on metallic nanotubes. The bandgap dependence is verified by analytical calculations, which also agree with preceding analyses for the limit of no curvature.
\end{abstract}

The electronic properties of single-walled carbon nano\-tubes (SWCNTs) have features making them suitable for a range of applications in quantum information processing and spintronics. Central to these potential applications is the energy dispersion. It has been predicted~\cite{Mint92,Hama92} and shown~\cite{Wild98,Odom98} that SWCNTs can be metallic as well as semiconducting. Transport experiments have demonstrated single-electron~\cite{Tans97,Post01} and field-effect~\cite{Tans98,Jave03} transistor action, among several other interesting physical effects. In these experiments gates have been applied to shift the electrostatic potential on the nano\-tubes; however, split gate structures can also be used to create electric fields. These fields are known to couple bands in the energy dispersion~\cite{Zhou01,Kim01,Vuko02,O'Ke02,Levi02,Li03}. If the split gates are deposited at different points along a SWCNT, artificial heterojunctions can be created and different types of quantum dot arrays can be produced. Gated structures of this kind have been fabricated for multi-walled carbon nano\-tubes~\cite{Robi03}.

The electronic properties of SWCNTs are normally determined by their chiral vector elements $(n,m)$ (see refs.~\cite{Wild98,Odom98} for a complete definition and experimental justification). Nano\-tubes satisfying $n-m=3p$, where $p$ is an integer, have become conventionally known as metallic nano\-tubes. However, most metallic nano\-tubes are in fact semi\-conductors due to curvature strain. These narrow-gap nano\-tubes have the bandgap (\textit{cf} ref.~\cite{Kane97})
\begin{equation}
  \label{e.1}
  E_\mathrm{g}^{(0)} = \frac{\hbar v_\mathrm{F} a_{\mathrm{cc}}}{8R^2}\cos 3\theta,
\end{equation}
where $v_\mathrm{F}\approx 7.25\cdot 10^{5}$m/s is the Fermi velocity of graphene, $a_{\mathrm{cc}}\approx 1.42$\AA~ is the carbon-carbon bond length, $R$ is the radius of the nano\-tube, and $\theta = \arctan[\sqrt{3}m/(2n+m)]$ is the chiral angle. The bandgaps in narrow-gap nano\-tubes have been observed experimentally~\cite{Ouya01}.

In this letter, we argue that narrow-gap nano\-tubes, in particular the zigzag ones ($\theta=0$), may be the most suitable for future transverse electric field experiments, including those using local gates of the kind described in ref.~\cite{Robi03}. This is due to the fact that an electric field causes a second-order coupling of the conduction and valence bands, and achivement of the required field amplitude is extemely challenging in practical experiments. Predictions suggest that the electric field across wide-gap semiconducting nano\-tubes must exceed a high threshold to modulate the bandgap~\cite{O'Ke02,Levi02}. In this letter, we show that narrow-gap nano\-tubes can be modulated without any threshold field.

\begin{figure}
  \onefigure{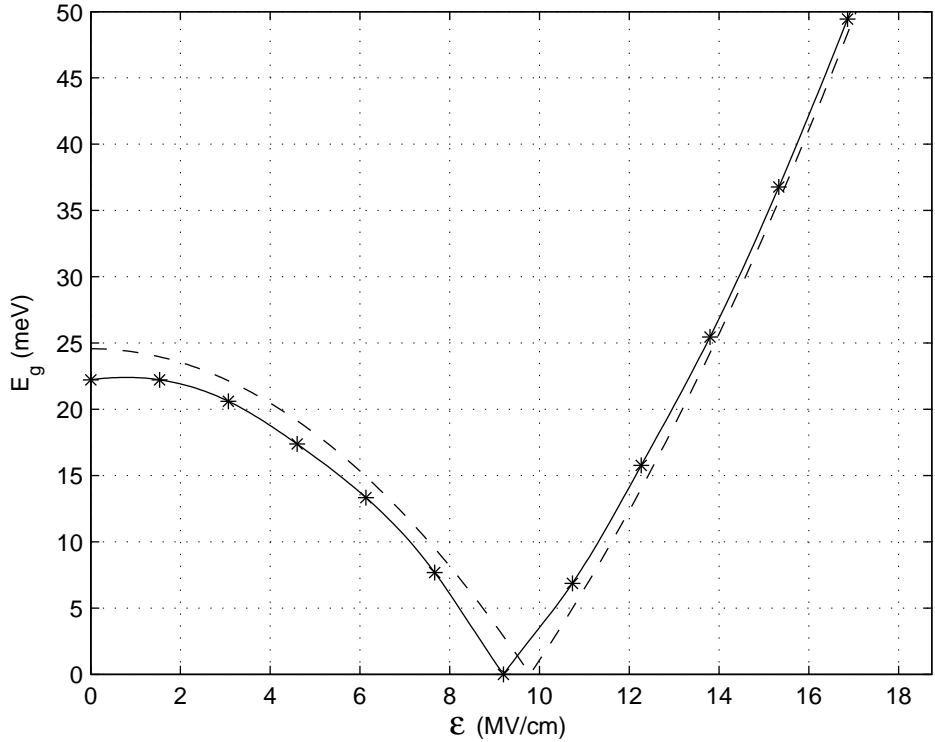}
  \caption{The bandgap of a typical narrow-gap zigzag nano\-tube, here $(15,0)$, is plotted as a function of transverse electric field amplitude across the nano\-tube. The solid curve represents a cubic spline interpolation of the numerical data points (where the modulus constraint was temporarily removed to deal with the discontinuity in the first derivative) and should be compared with the analytical prediction that is shown as the dashed curve. The initial gap is due to the curvature of the nano\-tube. As field amplitude increases, the bandgap is gradually suppressed to zero. The bandgap grows quadratically as the electric field amplitude is further increased.}
  \label{f.1}
\end{figure}

Many low-energy effects in carbon nano\-tubes can be predicted by a simple $\pi$-orbital model. However, the bandgaps in narrow-gap nano\-tubes owe their presence to hybridization between $\pi$- and $\sigma$-orbitals, and therefore we include all four $\pi$- and $\sigma$-orbitals in our model. The model is based on an orthogonal tight-binding (TB) approach with the following parameters: $E_s=-7.3$ eV, $E_p=0.0$ eV, $V_{ss\sigma}=-3.63$ eV, $V_{sp\sigma}=4.20$ eV, $V_{pp\sigma}=5.38$ eV, and $V_{pp\pi}=-2.24$ eV~\cite{Toma91}. We assume that the potential from the electric field varies smoothly on atomic scale, which enables it to be entered through the onsite energies. The bandgaps are obtained indirectly by calculating low-bias electron conductance through ideal nano\-tubes after applying a global electric field potential with various amplitudes to the nano\-tubes. The result for a $(15,0)$ zigzag SWCNT is shown as the solid curve in fig.~\ref{f.1}. 
At zero electric field the bandgap is approximately the same as in eq.~(\ref{e.1}). The observed quadratic dependence is a consequence of the selection rule $|p-p'|=1$, where $p$ and $p'$ denote the bands arising from the quantization of the circumferential wavevector in graphene~\cite{Kim01,Vuko02}. The selection rule can be viewed as an angular momentum conservation requirement, where only certain bands can be coupled by the momentum from the electric field. The conduction and valence bands have the same quantisation integer and do not satisfy this condition. Therefore, these bands are first coupled via second-order terms, which are quadratic in electric field amplitude.

The dependence of the bandgap on electric field can be estimated analytically by making use of a chiral transform~\cite{Levi02}. The Hamiltonian of our system can be expressed in three terms as
\begin{equation}
  \label{e.2}
  H = H_0 + H_{\mathrm{curv}} + H_{\mathrm{corr}},
\end{equation}
where $H_0$ is the first-order tight-binding Hamiltonian, $H_{\mathrm{curv}}$ is a curvature term, and $H_{\mathrm{corr}}$ is a second-order tight-binding correction term. The first term of the Hamiltonian is the same as the Weyl Hamiltonian for a massless relativistic spin-$1/2$ particle, that is
\begin{equation}
  \label{e.3}
  H_0 = -i\hbar v_\mathrm{F}~\vec \sigma \cdot\nabla + V,
\end{equation}
where $\vec \sigma$ is the Pauli matrix vector, $\vec q \equiv -i\vec{\partial} = -i(\partial_x,\partial_y)^T$ is the electron quasi-momentum operator, and the potential from the transverse electric field is
\begin{equation}
  \label{e.4}
  V = V_0 \cos \frac{y}{R}.
\end{equation}
The curvature term is
\begin{equation}
  \label{e.5}
  H_{\mathrm{curv}} = \lambda\hbar v_\mathrm{F}~\vec \sigma \cdot \hat{a}_c,
\end{equation}
where $\lambda\equiv a_{\mathrm{cc}}/16R^2$ and $\hat a_c = \sin 3\theta~\hat{x}+\cos 3\theta~\hat{y}$~\cite{Kane97}. $\theta$ is the chiral angle, 
which is $0$ for zigzag nano\-tubes and $\pi/6$ for armchair nano\-tubes. Finally, the correction term is
\begin{equation}
  \label{e.6}
  H_{\mathrm{corr}} = \eta\hbar 
v_\mathrm{F}~e^{i\frac{3\theta}{2}\sigma_z}[2\sigma_x\partial_x\partial_y+\sigma_y(\partial_x^2-\partial_y^2)]e^{-i\frac{3\theta}{2}\sigma_z},
\end{equation}
where $\eta\equiv a_{\mathrm{cc}}/4$. The rotation, $R_z(3\theta)$, makes this expression valid for all carbon nano\-tubes, including chiral nano\-tubes. There is no explicit $x$-dependence in the Hamiltonian. Therefore, the wavefunction can be separated, $\Psi(x,y)=\psi(x)\phi(y)$, where the boundary condition in the circumferential direction is periodic for the narrow-gap nano\-tubes covered in this paper. The circumferential $y$-dependence in the potential can be eliminated by the chiral gauge transformation~\cite{Levi02}
\begin{subequations}
  \label{e.7}
  \begin{eqnarray}
    \phi(y) &=& e^{-i\sigma_y\zeta(y)}~\tilde \phi(y),
    \\
    \zeta(y) &\equiv& \frac{V_0}{\hbar v_\mathrm{F}}R\sin\frac{y}{R}.
  \end{eqnarray}
\end{subequations}
This transformation leaves the periodic boundary condition of the wavefunction $\phi(y)$ unchanged, but leads to the new Hamiltonian
\begin{equation}
  \label{e.8}
  \tilde H = e^{i\sigma_y\zeta(y)}~H~e^{-i\sigma_y\zeta(y)}.
\end{equation}
The conduction and valence bands of a narrow-gap nano\-tube have the quantisation condition, $q_y=0$. Therefore, we can treat the transverse electric 
field as a perturbation, where the unperturbed wavefunction $\tilde\phi$ is constant. By integrating over the $y$-coordinate in the envelope function equation, re-aligning the dispersion to $q_x=0$, and approximating the Hamiltonian to first-order, we obtain the following one-dimensional effective Hamitonian:
\begin{equation}
  \label{e.9}
  \tilde H_{\mathrm{1D}} = \hbar v_\mathrm{F}\cos 3\theta~\left[\lambda+\frac{\eta}{2}\left(\frac{V_0}{\hbar v_\mathrm{F}}\right)^2\right]\sigma_y-i\hbar v_\mathrm{F}J_0\left(\frac{2V_0R}{\hbar v_\mathrm{F}}\right)\sigma_x\partial_x.
\end{equation}
This Hamiltonian has the form of a semiconducting tight-binding Hamiltonian. However, the usual group velocity has an extra Bessel function factor, which predicts slower low-energy electrons in the presence of a transverse electric field. This effect is so strong that the electron velocity undergoes a sign-reversal at the first Bessel function node~\cite{Levi02}. Substituting $V_0 = \Leps R$, where $\Leps$ is the electric field amplitude, we find that the critical field is $\Leps_\mathrm{c} = 1.2024 \hbar v_\mathrm{F}/R^2$. Above this field the direct bandgap moves away from $k=0$, and our model breaks down. Below the critical field, on the other hand, we can use eq.~(\ref{e.9}) to calculate the bandgap, yielding
\begin{equation}
  \label{e.10}
  E_\mathrm{g}(\Leps) = \frac{\hbar 
    v_\mathrm{F}a_{\mathrm{cc}}}{4}\cos 3\theta\left|\frac{1}{2R^2}-\left(\frac{R}{\hbar 
    v_\mathrm{F}}\right)^2\Leps^2\right|.
\end{equation}
This function is plotted as the dashed line in fig.~\ref{f.1} and exhibits the same characteristics as the numerical results. There is no anti-crossing in our model; instead, we can calculate the electric field $\Leps^*$, at which the bandgap closes for narrow-gap chiral or zigzag nano\-tubes. Identifying $E_\mathrm{g}(0)=E_\mathrm{g}^{(0)}$ [\textit{cf} eq.~(\ref{e.1})] gives
\begin{equation}
  \label{e.11}
  \Leps^* = \frac{4\sqrt{2}E_\mathrm{g}^{(0)}}{a_{\mathrm{cc}}\cos 3\theta} = \frac{\hbar v_\mathrm{F}}{\sqrt{2}R^2}.
\end{equation}
Eqs.~(\ref{e.10}) and (\ref{e.11}) reveal that the bandgaps of narrow-gap nano\-tubes with larger radii respond more strongly to a transverse electric field. This is due to the increased potential difference across these nano\-tubes. Larger nano\-tubes have also smaller initial bandgaps [see eq.~(\ref{e.1})] because their curvature strains are less. In the limit of no curvature eq.~(\ref{e.10}) simplifies to the bandgap dependence reported in refs.~\cite{Levi02,Li03}.

\begin{figure}
  \onefigure{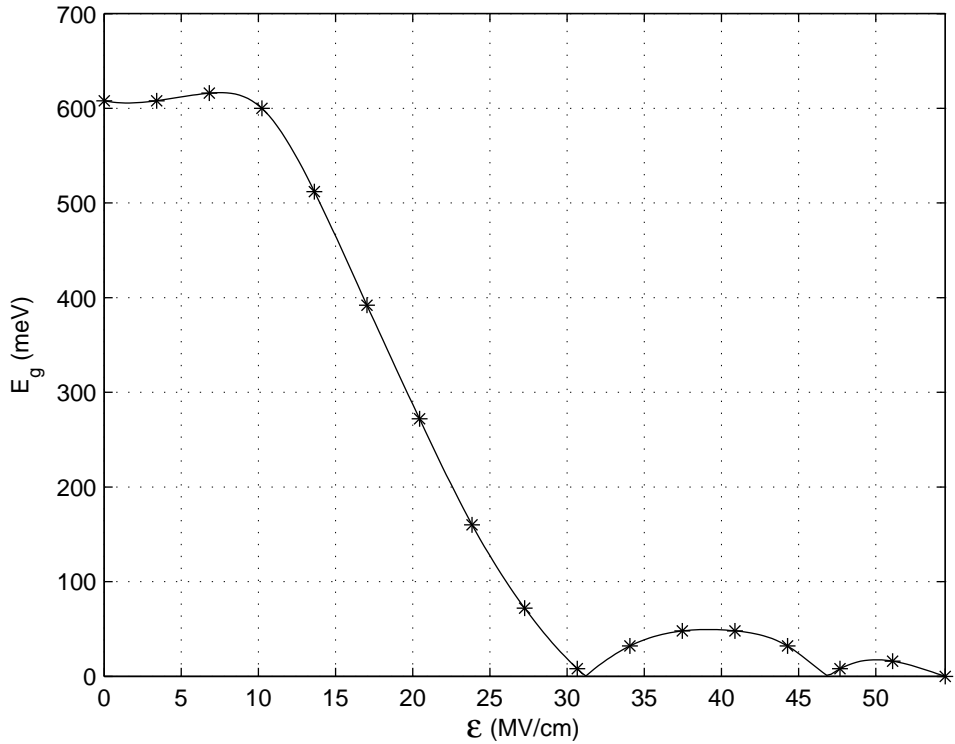}
  \caption{The bandgap of a typical wide-gap zigzag nano\-tube, here $(17,0)$, is plotted as a function of transverse electric field amplitude across the nano\-tube. The solid curve represents a cubic spline interpolation of the numerical data points (where the modulus constraint was temporarily removed to deal with the discontinuities in the first derivative). The electric field amplitude needs to exceed a threshold amplitude, in this case about $10$ MeV/cm, before the bandgap becomes noticeably modulated. Thereafter, the bandgap decreases almost linearly to zero and then continues to succesively open and close.}
  \label{f.2}
\end{figure}

The main challenge in utilizing a transverse electric field for bandgap modulations stems from the required field amplitudes. The electric field generated by a split gate in an actual experiment is partly dielectrically screened. The dielectric constant in narrow-gap nano\-tubes is of order $4-5$ in the transverse direction~\cite{Bene95,Levi02,Li03,Krcm03}. An electric field amplitude of $5$~MeV/cm between the split gates is sufficient to close the bandgaps in very large ($R\simeq 18$~\AA) narrow-gap nano\-tubes. The same field should also be able to make smaller bandgap modulations in smaller nano\-tubes. Wide-gap semiconducting nano\-tubes behave rather differently and require that the electric field overcome a high threshold in order to affect the bandgap~\cite{O'Ke02,Levi02}. Fig.~\ref{f.2} confirms this from our modelling of wide-gap semiconducting nano\-tubes, and shows that at sufficiently high field there can be multiple closures of the bandgap.
The bandgap oscillations also occur in narrow-gap nanotubes at high fields, $\Leps > \Leps_\mathrm{c}$. The threshold field is approximately $\Leps_{\mathrm{th}}\approx 0.6215 \hbar v_\mathrm{F}/R^2$~\cite{Levi02} which is about $12\%$ less than $\Leps^*$ (\textit{cf} the threshold field in fig.~\ref{f.2} with the closure field in fig.~\ref{f.1}). Since a threshold field in wide-gap nanotubes is approximately the same as the closure field in narrow-gap nanotubes with the same radii, we expect the required closure field to be significantly less in the narrow-gap nanotubes. Because of the difference in required electric field amplitudes, narrow-gap nano\-tubes are likely to be the most suitable for use in field-modulated electron transport experiments.

\acknowledgments
This research is part of the QIP IRC (GR/S82176/01) and was supported through the Foresight LINK Award Nanoelectronics at the Quantum Edge by EPSRC (GR/R660029/01) and Hitachi Europe Ltd. GADB thanks EPSRC for a Professorial Research Fellowship (GR/S15808/01) and JHJ acknowledges support from the UK MOD.


\begin{thebibliography}{0}

\bibitem{Mint92}
  \Name{Mintmire J. W., Dunlap B. I. \and White C. T.}
  \REVIEW{Phys. Rev. Lett.}{68}{1992}{631}.

\bibitem{Hama92}
  \Name{Hamada N., Sawada S.-i. \and Oshiyama A.}
  \REVIEW{Phys. Rev. Lett.}{68}{1992}{1579}.

\bibitem{Wild98}
  \Name{Wild\"oer J. W. G., Vanema L. C., Rinzler A. G., Smalley R. E. \and Dekker C.}
  \REVIEW{Nature}{391}{1998}{59}.

\bibitem{Odom98}
  \Name{Odom T. W., Huang J.-L., Kim P. \and Lieber C. M.}
  \REVIEW{Nature}{391}{1998}{62}.

\bibitem{Tans97}
  \Name{Tans S. J., Devoret M. H., Dai H. J., Thess A., Smalley R. E., Geerligs L. J. \and Dekker C.}
  \REVIEW{Nature}{386}{1997}{474}.

\bibitem{Post01}
  \Name{Postma H. W. Ch., Teepen T., Yao Z., Grifoni M. \and Dekker C.}
  \REVIEW{Science}{293}{2001}{76}.

\bibitem{Tans98}
  \Name{Tans S. J., Verschueren A. R. M. \and Dekker C.}
  \REVIEW{Nature}{393}{1998}{49}.

\bibitem{Jave03}
  \Name{Javey A., Guo J., Wang Q., Lundstrom M. \and Dai H. J.}
  \REVIEW{Nature}{424}{2003}{654}.

\bibitem{Zhou01}
  \Name{Zhou X., Chen H. \and Zhong-can O.-Y}
  \REVIEW{J. Phys. Cond. Mat.}{13}{2001}{L635}.

\bibitem{Kim01}
  \Name{Kim Y.-H. \and Chang K. J.}
  \REVIEW{Phys. Rev. B}{64}{2001}{153404}.

\bibitem{Vuko02}
  \Name{Vukovi\'{c} T., Milo\v{s}evi\'{c} I. \and Damnjanovi\'{c} M.}
  \REVIEW{Phys. Rev. B}{65}{2002}{045418}.

\bibitem{O'Ke02}
  \Name{O'Keeffe J., Wei C. \and Cho K.}
  \REVIEW{Appl. Phys. Lett.}{80}{2002}{676}.

\bibitem{Levi02}
  \Name{Levitov L. S. \and Novikov D. S.} in
  \Book{Quantum Phenomena in Mesoscopic Systems, Proceedings of the International School of Physics ``Enrico Fermi'', Course CLI}
  \Editor{Altshuler B., Tagliacozzo A. \and Tognetti V.}
  \Publ{IOS Press, Amsterdam}
  \Year{2003}
  \Pages{357}{373}.

\bibitem{Li03}
  \Name{Li Y., Rotkin S. V. \and Ravaioli U.}
  \REVIEW{Nano Lett.}{3}{2003}{183}.

\bibitem{Robi03}
  \Name{Robinson L. A. W., Lee S.-B., Teo K. B. K., Chhowalla M., Amaratunga G. A. J., Milne W. I., Williams D. A., Hasko D. G. \and Ahmed H.}
  \REVIEW{Nanotechnology}{14}{2003}{290};
  \REVIEW{Microelectron. Eng.}{67}{2003}{615}.

\bibitem{Kane97}
  \Name{Kane C. L. \and Mele E. J.}
  \REVIEW{Phys. Rev. Lett.}{78}{1997}{1932}.

\bibitem{Ouya01}
  \Name{Ouyang M., Huang J.-L., Cheung C. L. \and Lieber C. M.}
  \REVIEW{Science}{292}{2001}{702}.

\bibitem{Toma91}
  \Name{Tomanek D. \and Schluter M. A.}
  \REVIEW{Phys. Rev. Lett.}{67}{1991}{2331}.

\bibitem{Bene95}
  \Name{Benedict L. X., Louie S. G. \and Cohen M. L.}
  \REVIEW{Phys. Rev. B}{52}{1995}{8541}.

\bibitem{Krcm03}
  \Name{Kr\v{c}mar M., Saslow W. M. \and Zangwill A.}
  \REVIEW{J. Appl. Phys.}{93}{2003}{3495}.

\end{thebibliography}
\end{document}